\documentstyle[prl,twocolumn,aps,graphicx]{revtex}
\begin{document}
\bibliographystyle{prsty}

 \newcommand{\mytitle}[1]{
 \twocolumn[\hsize\textwidth\columnwidth\hsize
 \csname@twocolumnfalse\endcsname #1 \vspace{5mm}]}

\mytitle{
\title{Anomalous Scaling of the SO(8) Symmetric Phases in the Two-Leg 
Ladder}
\author{Hsiu-Hau Lin}
\address{Department of Physics, National Tsing-Hua University, 
Hsinchu 300,Taiwan}
\date{\today}
\maketitle

\begin{abstract}
We carry out a complete analytical stability study for the SO(8) 
symmetric phases in the weakly-interacting two-leg ladder. It is 
shown that the SO(8) symmetry is robust under generic perturbations. 
Since there are no fixed points in the one-loop renormalization group 
equations, the conventional classification of relevant and irrelevant 
perturbations fails in this case. A new classification is defined and 
explained in detail. It leads to the anomalous scaling in the ratios 
of excitation gaps and an universal exponent $1/3$ is extracted. This 
new method is also applied to the well studied 2D Ising model and 
similar exponent is calculated.
\end{abstract}
}

{\it Introduction.---}
Ladder materials have attracted lots of attentions in the past decade 
because of both theoretical and experimental interest \cite{Dagotto96}.
In particular, lots of efforts were focused on the simplest two-leg 
ladder \cite{Balents96,Lin98,Noack94,Noack95,Dagotto92,Sano96,Troyer96,Vojta99,Kim99,Oitmaa99}.
Without doping, there is one electron per site on average (usually 
referred as ``half-filled''). Because of the Coulomb repulsion, 
charge excitations acquire a gap which makes the two-leg ladder a 
Mott's insulator. Besides, due to the tendency for singlet bond 
formation across the rungs of the ladder, spin-liquid behavior is 
expected. Both numerical and analytical approaches support that the 
ground state at half filling is in the Mott-insulating spin-liquid 
phase \cite{Balents96,Lin98,Noack94,Noack95}.

In general, one expects that charge and spin gaps could be rather 
different. Surprisingly, a complete degeneracy of charge, spin and 
single particle gaps emerges in the asymptotic weak coupling limit. 
Based on the one-loop renormalization group (RG) analysis, the 
effective low-energy theory for the two-leg ladder is described by 
the exactly soluble Gross-Neveu model with a global SO(8) symmetry. 
Other than the exact degeneracy of excitation gaps, many interesting 
results can be drawn from the Bethe Ansatz solution. There have been 
critics \cite{Azaria98,Emery99} about the unexpected restoration of the 
SO(8) symmetry in weak coupling, concerning about the stability of the 
symmetry under generic perturbations and the limitation of the RG 
equations derived in weak coupling. It was argued that the SO(8) 
symmetry derived by perturbative calculations is fragile because the 
system eventually flows toward the fixed point in strong coupling.

In this Letter, we address both points by performing a complete an
analytical study for the stability in the vicinity of the SO(8)
symmetric phases. It should be emphasized that this beautiful 
symmetry restoration in weak coupling does {\it not} implies that the 
fixed point in strong coupling is SO(8) symmetric at all. On the 
contrary, the symmetry reflects the local topology of RG flows near 
the trivial Fermi liquid fixed point which can be safely described by 
perturbative calculations. Indeed it is rather obvious that, when the 
interaction $U$ is much larger than the bandwidth $t$, the charge gap 
is much larger than the spin gap. Thus, the symmetry in the strong 
coupling is completely destroyed. As to the stability of the symmetry 
under generic perturbations, the complete stability check shows that 
the SO(8) symmetry is robust in weak coupling. Even though some 
couplings are relevant according to the conventional classification, 
they do not destroy the symmetry but only give rise to anomalous 
corrections which scale as $(U/t)^{1/3}$. A new classification of 
``relevant'' and ``irrelevant'' perturbations is necessary here. 
Finally, we apply the new method to the well-known 2D Ising model and 
show that the anomalous scaling behavior near the critical point can 
be very general, as long as more than two relevant couplings are 
present.

{\it The SO(8) symmetric rays.---}
For the two-leg ladder at half filling, the number of possible 
interactions is greatly reduced to nine in weak coupling. Within the 
one-loop RG calculations, these nine couplings $g_{i}$ are described 
by a set of coupled first-order differential equations,
\begin{equation}
\frac{dg_{i}}{dl} = A^{i}_{jk} g_{j} g_{k},
\label{RGFlow}
\end{equation}
where $A^{i}_{jk}$ are nine $9 \times 9$ constant matrices. These 
matrices can be found by rewriting the RG equations given in 
Ref.\cite{Lin98}. For a generic interacting Hamiltonian, the bare 
values for these nine couplings can be straightforwardly determined. 
However, it is generally very difficult to obtain the solution for 
these coupled flow equations in analytical form.

Simple analytical solutions emerge if the interactions are chosen in a 
specific way. These special solutions are later referred as 
``symmetric rays''. Suppose the bare couplings of the specific 
interacting Hamiltonian are $ g_{i}(0) = r_{i} g(0)$, where $g(0) = 
(U/t) \ll 1$ is small while $r_{i}$ are order one constants which 
satisfy the algebraic constraint,
\begin{equation}
r_{i} = A^{i}_{jk} r_{j} r _{k}.	
\end{equation}
It is straightforward to show that the ratios between couplings 
remain the same and the nine complicated equations reduce to single 
one, $\dot{g} = g^{2}$ ! The solution is $g(l)=1/(l_{d}-l)$, where 
the divergent length scale $l_{d} = (t/U)$. Of course, one should 
keep in mind that the solution $g(l)$ is only valid when it does not 
flow out of the weak coupling regime. These special Hamiltonians, 
whose couplings are described by these symmetric rays, turn out to be 
SO(8) symmetric. For the two-leg ladder, four different phases (named 
as D-Mott, S-Mott, CDW and SP in Ref.\cite{Lin98}) are of the central 
concerns.

It was shown previously that the two-leg ladder in weak coupling {\it 
always} scales into one of the four different symmetric phases 
\cite{Lin98}. However, this numerical approach was criticized that the 
SO(8) symmetric phases might have instabilities which happen not to be 
tackled by the limited types of interactions considered in the 
numerical study. To make up the fissure, a complete stability check 
near the SO(8) symmetric rays is desirable.

{\it Stability analysis.---}
To describe the RG flows in the vicinity of the SO(8) symmetric rays, 
it is sufficient to consider the linearized version of Eq.~(\ref{RGFlow}). 
For a generic interaction, the couplings are separated into symmetric 
and asymmetric parts, $g_{i}(l) = r_{i} g(l) + \Delta g_{i}(l)$. In 
the vicinity of the symmetric rays, the deviations are small, $\Delta 
g_{i}(l) \ll g(l)$. Keeping the leading order term, the linearized RG 
equations are
\begin{equation}
\frac{d (\Delta g_{i})}{dl} = \frac{B_{ij}}{(l_{d}-l)} \Delta g_{i},
\end{equation}
where $B_{ij} = 2 A^{i}_{jk} r_{k}$. The matrix $B_{ij}$ can be 
brought into diagonal form by a linear transformation. As a 
consequence, the RG equations decouple into nine independent ones,
\begin{equation}
\frac{d (\delta g_{i})}{dl} = \frac{\lambda_{i}}{(l_{d}-l)} \delta g_{i},
\label{LinearRG}
\end{equation}
where $\delta g_{i}$ are couplings after the linear transformation 
and $\lambda_{i}$ are the eigenvalues of the matrix $B_{ij}$. Although 
the matrix $B_{ij}$ are different for each SO(8) symmetric rays, the 
eigenvalues are identically the same
\begin{equation}
\lambda_{i} = 2, \frac23, \frac23, \frac23, 
-\frac13, -\frac13, -\frac13, -\frac13, -\frac13.
\end{equation}
This coincidence implies that the results of the stability check only 
rely on the symmetry group but not on the details of the phases 
\cite{Konik00}.

So far, we have a single equation, $\dot{g}=g^{2}$, describing the
renormalization along the symmetric rays and nine for deviations from
the rays as in Eq.~(\ref{LinearRG}).  Apparently, there must be one
redundant equation among them because the original number of equations
is only nine.  It doesn't take long to find out that it corresponds to
the flow equation with largest eigenvalue $\lambda=2$. This 
corresponds to a trivial case that all couplings are shifted along the 
symmetric ray, i.e. $\delta g_{i} = r_{i} \delta g$. Since the flow 
along the ray is described by $\dot{g}=g^{2}$, linearization leads 
to $\delta \dot{g}= 2g(l) \delta g$. Since we only need eight 
equations to describe the deviations from the symmetric rays, the 
$\lambda=2$ equation is only an artifact and should be ignored.

The other eight eigenvalues describe how the flow goes once the bare 
couplings are off the ray. Starting from the bare values $\delta 
g_{i}(0) = \delta r_{i} g(0)$, where $\delta r_{i} \ll r_{i}$. The 
solutions of Eq.~(\ref{LinearRG}) are
\begin{equation}
\delta g_{i}(l) = \frac{\delta r_{i}}{(l_{d}-l)^{\lambda_{i}}} 
\left(\frac{U}{t} \right)^{1-\lambda_{i}}.	
\label{Deviations}
\end{equation}
According to conventional classification \cite{Goldenfeld92}, for 
$\lambda_{i}<0$ the deviations diminish under RG transformations as 
shown in FIG. 1(a) and thus are classified as irrelevant couplings. 
For $\lambda_{i}>0$, any small deviations get enhanced and the flow 
is pushed away from the symmetric rays as in FIG. 1(b) and 1(c). These 
are classified as relevant couplings. The conventional wisdom tells us 
that there are three relevant couplings ($\lambda_{i} = 2/3$) and five 
irrelevant ones ($\lambda_{i}=-1/3$). Since a generic interaction in 
principle could generate asymmetric deviations in all couplings, one 
might rush to the incorrect conclusion that the SO(8) symmetry is not 
stable. However, the first-glance guess is wrong because the 
conventional classification is based on the perturbative analysis 
near a $fixed point$ while we are dealing with running symmetric rays. 
A new set of rules to identify relevant perturbations is in order.

\begin{figure}[htb]
\centering
\includegraphics[height=3cm]{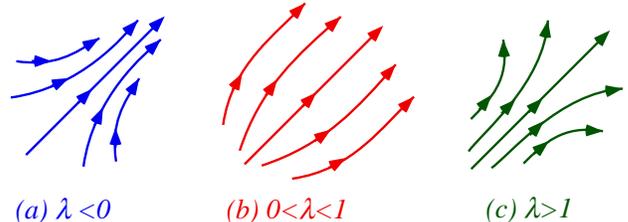}
\caption{The topology of RG flows near the symmetric ray with (a) 
$\lambda<0$, (b) $0<\lambda<1$ and (c) $\lambda>1$. It is clear that 
the coupling is irrelevant for $\lambda<0$ and relevant for 
$\lambda>1$. The RG flow is more subtle for $0<\lambda<1$. In this 
case, although the deviation from the symmetric ray is growing, the 
slope remains the same as the ray.}
\end{figure}

The crucial criterion is whether the deviations $\delta g_{i}(l)$ grow
larger than the symmetric coupling $g(l)=(l_{d}-l)^{-1}$.  Just
growing larger than the bare values under RG transformations is not
qualified as a relevant perturbation. This subtle but important 
difference is best illustrated by calculating gap functions using 
scaling arguments. If the degeneracy of excitation gaps is 
maintained, the SO(8) symmetry is robust and vice versa.

{\it Anomalous scaling.---}
Since the RG equations are only valid in weak coupling, we cut off 
the RG procedure when the symmetric coupling $g(l_{c})=1$. At this 
cutoff length scale $l_{c}$, the deviations in Eq.~(\ref{Deviations}) 
are
\begin{equation}
\delta g_{i} (l_{c}) \sim \left( \frac{U}{t}\right)^{1-\lambda_{i}}.
\end{equation}
For $\lambda_{i}>1$, the deviations are larger than the symmetric 
coupling and should be classified as ``relevant''. As long as 
$\lambda_{i}<1$, the deviations at the cutoff length scale are still 
vanishingly small and should be viewed as ``irrelevant''. This new 
classification is different from the conventional one for 
$0<\lambda_{i}<1$. We would see clearly soon that the new 
classification is appropriate for stability check near a running 
symmetric ray. The grey regime $0<\lambda_{i}<1$ between the new and 
conventional rules gives rise to anomalous scaling exponents.

Under RG transformations, the gap functions scale like,
\begin{equation}
\Delta_{i} \left[ g(0), \delta g_{i}(0) \right] =
e^{-l_{c}} \Delta_{i} \left[ 1, \delta g_{i}(l_{c}) \right],
\end{equation}
where $\delta g_{i}(l_{c})$ are at most of order $(U/t)^{1/3}$. The 
key point is that, although some perturbations are enhanced to 
order $(U/t)^{1/3}$, which is larger than the bare order $U/t$ 
values, they are still small at the cutoff length scale. The 
effective Hamiltonian can be separated into two parts $H = H_{0} + 
\delta H$ -- the SO(8) symmetric and asymmetric parts. Since the 
asymmetric part is of order $(U/t)^{1/3}$, the changes of the gaps 
would be of the same order by standard perturbation theory. Without 
the deviations, the SO(8) symmetry guarantees the exact degeneracy of 
all gaps, i.e. $\Delta[1,0] = \Delta$. The presence of perturbations 
modifies the gap functions,
\begin{equation}
\Delta_{i}\left[1,\delta g_{i}(l_{c}) \right] = \Delta \left[
1+c_{i} \left(\frac{U}{t}\right)^{\frac13} + \ldots \right].
\end{equation}
It is clear that, in the weak coupling limit $U/t \to 0$, the 
degeneracy of all gaps is recovered. It implies that the SO(8) 
symmetry is indeed robust under generic perturbations.

\begin{figure}[htb]
\centering
\includegraphics[height=4cm]{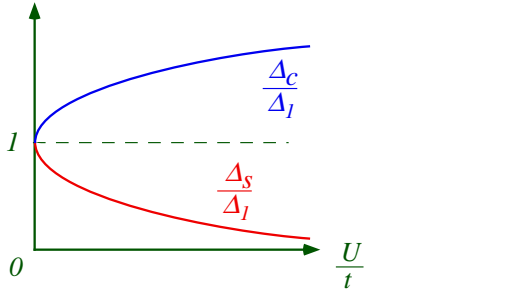}
\caption{Ratios of charge, spin and single particle gaps plotted 
versus the interaction strength $U/t$. The ratios approach unity in 
the asymptotic $ U/t \to 0$ limit with anomalous corrections of 
order $(U/t)^{1/3}$.}
\end{figure}

The anomalous scaling exponent $1/3$ is clearly seen in the ratios 
between charge, spin and single particle gaps (see FIG. 2),
\begin{equation}
\frac{\Delta_{i}}{\Delta_{j}} \approx 1 + c_{ij} \left( \frac{U}{t} 
\right)^{\frac13}.
\end{equation}
This simple exponent is rather fascinating because the RG equations we 
started from are very messy. But, despite of which SO(8) symmetric 
phases the system flows into, the exponent of the gap function 
corrections is universally equal to $1-\lambda_{i} = 1/3$!

\begin{figure}
\centering
\includegraphics[height=3.5cm]{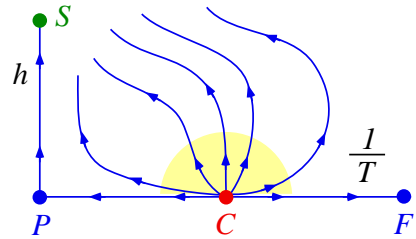}
\caption{The RG flows near the critical point in the 2D Ising model. 
Notice that the local topology of flows near the critical point is 
similar to FIG. 1(b).}
\end{figure}

{\it 2D Ising model.---}
The anomalous scaling discussed here is not limited to the particular 
two-leg ladder at all. As long as more than two relevant couplings 
(according to the conventional classification) are present, this 
interesting phenomena shows up somewhere. Here we use the 2D Ising 
model as another example. The RG flow diagram for the model is shown 
in FIG. 3. There are two relevant perturbations near the critical 
point -- the magnetic field $h$ and the temperature deviation from 
the critical point $t$. It is well known that the magnetization 
$m[h,t]$ scales differently with respect to these two relevant 
perturbations,
\begin{eqnarray}
m[u,0] &\sim& u^{1/\delta},
\\
m[0,u] &\sim& u^{\beta}.
\end{eqnarray}
Here $1/\delta=1/15$ and $\beta=1/8$. Although both are relevant, the 
magnetic field grow faster than the later under RG transformations. We 
would find out that the scaling of magnetization $m[h,t]$ in the 
presence of both perturbations looks similar to the scaling behavior 
with only non-zero magnetic field. Suppose now we start from a bare 
coupling in the vicinity of the critical point, $(h,t) = u 
(\cos\theta, \sin\theta)$ with $0<\theta<\pi/2$. The scaling function 
of the magnetization is $m[h,t] = h^{1/\delta} F(t^{\Delta}/h)$ with 
the exponent $\Delta = \beta \delta = 15/8 >1$. Since the argument 
inside the scaling function $t^{\Delta}/h \sim u^{\Delta-1}$ is small, 
we can expand the function around zero. After a bit algebra, the 
magnetization is
\begin{equation}
\frac{m[u\cos\theta, u\sin\theta]}{m[u,0]} \approx 1 + c(\theta) u^{7/8}.
\end{equation}
Notice that the exponent comes from $\Delta-1 = 7/8$. It is clear 
that, when close to the critical point $u \to 0$, the 
magnetization scales as if the temperature deviation is not present 
at all. In this sense, the temperature deviation $t$ should be called 
``irrelevant''. However, it does lead to a non-trivial correction 
with anomalous exponent $7/8$.

In summary, we have shown analytically that the SO(8) symmetric phases 
are stable in weak coupling. Since we are dealing with symmetric rays 
but not fixed points, a new classification is defined and explained 
in detail. Corrections to the SO(8) symmetry at finite but still weak 
couplings acquire a non-trivial exponent $1/3$ in the ratios of 
excitation gaps.

We thank Leon Balents and Matthew Fisher for useful discussions. This 
work is supported by National Science Council in Taiwan under grant 
number NSC 89-2112-M-007-101.

\end{document}